\documentclass[twocolumn,showpacs,preprintnumbers]{revtex4}
\usepackage{graphicx}

\begin{document}

\title{
Coupled dynamics of an atom and an optomechanical cavity}
\author{ X. X. Yi, H. Y. Sun, and L. C. Wang}

\affiliation{School of Physics and Optoelectronic Technology, Dalian
University of Technology, Dalian 116024, China }

\begin{abstract}
We consider the motion of the end mirror of a cavity inside which a
two-level atom trapped. The fast vibrating mirror induces nonlinear
couplings between the cavity field and the atom. We analyze this
optical effect by showing the population of the atom in its internal
degrees of freedom as a function of time. On the other side, fast
atom-field variables result in an additional potential for the
atomic center-of-mass motion and the mirror vibration, leading to
entanglement in the motion and the vibration. The entanglement has
been numerically simulated and discussed.
\end{abstract}

 \maketitle

\section{introduction}

Optomechanical cavities
\cite{cohadon99,gigan06,kleckner06,arcizet06,schliesser06,
corbitt07} in which electromagnetic degrees of freedom couple to the
mechanical motion of mesoscopic or macroscopic mirrors are promising
candidates for studying the transition of a macroscopic degree of
freedom from the classical to the quantum regime. These systems also
offer the prospect of technological use, for example the single
molecule detection\cite{yang06}, the gravitational wave
detection\cite{courty03} and the possible new quantum information
processing devices\cite{mancini03}. In these
radiation-pressure-driven devices, the number of photons trapped
inside the cavity is a key variable, since the radiation pressure on
the mirror is proportional to the photon number. Several strategies
to increase this optomechanical coupling have been proposed. In
Ref.\cite{meiser06}, the authors have developed a model to describe
the coupled motion of a cavity end mirror and cold atoms trapped
inside the cavity. It was shown that the atoms can from a
distributed Bragg mirror with high reflectivity\cite{note1}, leading
to a superstrong coupling regime for the cavity quantum
electrodynamics(CQED) system. In this proposal, the atoms inside the
cavity are assuming an initial Bose-Einstein condensate.  This
requirement for the atom medium was lifted in Ref.\cite{ian08},
where the low-energy collective excitations of the atoms were used
to enhance the coupling between the mirror and the cavity field.

While the superstrong coupling regime has not yet been reached in
experiments, a regime where the mechanical oscillation frequency is
larger than the cavity linewidth has recently been
observed\cite{schliesser08}. In this quantum regime, it is
interesting to explore entanglement shared between mechanical
(macroscopic) and microscopic degrees of freedom. This is one of our
goals in this paper. In fact, possibilities to entangle the
oscillatory motion of a cavity macromirror with the electromagnetic
field in the cavity have been explored in various
approaches\cite{mancini02,vitali07}, steady state entanglement in
the mechanical vibrations of two macroscopic membranes has been
studied\cite{hartmann08}.

In this paper, we consider a Fabry-P\'{e}rot (F-P) cavity with a
moving end mirror, which is allowed to move under the effect of
radiation pressure(see Fig.\ref{fig1}). An atom is trapped in the
standing-wave light field of the cavity with frequency $\omega$. We
shall show that the mechanical vibrations of the end mirror and the
atomic center-of-mass(COM) motion can be entangled. Due to the
vibration dependent coupling between the atom and the cavity field,
the CQED system induces interesting phenomena worth investigating.
In contrast to earlier works, our study considers only  a single
atom inside the cavity, the atomic COM motion and the position
dependent atom-to-field couplings may thus be taken into account due
to the simplicity of
 systems involved.

\section{model}

\begin{figure}
\includegraphics*[width=0.6\columnwidth,
height=0.5\columnwidth]{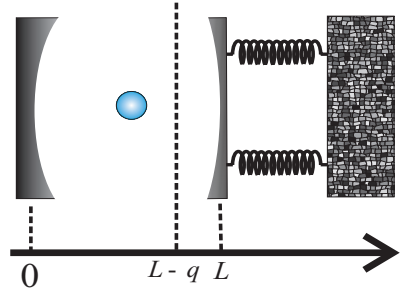}\caption{(Color
online)Schematic illustration of an atom in a cavity with a movable
end mirror.} \label{fig1}
\end{figure}

 The system under consideration is shown schematically in
 Fig.\ref{fig1}.
 An two-level atom with Rabi frequency $\Omega$ is
 trapped in the standing-wave light field of a F-P cavity with one
 of its end mirrors allowed to move and subject to a harmonic
 restoring force, $q$ being the displacement of the mirror from its
 rest position. The cavity length is $L$ and the mass of the movable
 mirror is $m$. Although radiation pressure excites several
 mechanical degrees of freedom, coupling among the different
 vibrational modes can typically be neglected\cite{vitali07,
 hartmann08,pinard99}. We also assume that the electromagnetic field
 frequency $\omega$ follows adiabatically the mirror vibration,
 hence  the frequency of the cavity field $\omega$ is simply
 parameterized by the mirror position/vibration  $q$. The Hamiltonian of such a
 system is then\cite{bhattacharya07},
\begin{eqnarray}
H&=&\hbar\omega a^{\dag}a+\frac{p^2}{2m}+\frac 1
2m\omega_m^2q^2+\frac{P^2}{2M}+\frac{\hbar\Omega}2\sigma_z\nonumber\\
&-&\hbar\xi a^{\dag}aq +\hbar
g\sin(kQ)(a^{\dag}\sigma^-+h.c.),\label{h1}
\end{eqnarray}
where $\omega_m$ denotes the frequency of the mirror vibration, $M $
$(m)$ is the mass of the atom (mirror), and $P$  $(p)$ stands for
the momentum of the atom (mirror). The atom-field coupling constant
$\hbar g\sin(kQ)$ depends on the  atom position  $Q$ as well as the
vibration $q$ of the mirror  through $k=\omega_{eff}/c$, where
$\omega_{eff}=\omega-\xi q$\cite{bhattacharya07}, and
$\xi=\omega/L$. The most interesting part of the dynamics arises
from the coupling term $\hbar g\sin(kQ)(a^{\dag}\sigma^-+h.c.)$,
describing interactions among the atom, the atomic center-of-mass
motion, the cavity field and the vibration of the mirror.  We shall
analyze its effects in two limiting cases. (a)Slow atomic COM
motion: in this regime, we can safely  ignore the COM motion of the
atom, describing a very cold atom or an atom at a fixed position
inside the cavity, and (b) the cavity field is large detuned from
the atomic transitions, such that the atomic center-of-mass motion
and the vibration of the cavity mirror is slow. In this situation,
the center-of-mass motion and the mirror vibration can be  included
in the analysis. We shall explore the entanglement created among the
motion and vibration.

\section{slow atomic center-of-mass motion}
In this section we study the dynamics of the atom-cavity system when
the atomic center-of-mass can be safely ignore. From Eq.(\ref{h1})
 we find that the atom-cavity coupling depends on the mirror
 vibration through $\sin(kQ)$. In general, the dynamics governed by
 this coupling can not be  analytically solved. Hence we shall
 consider two limiting cases, which may loss some physics with the
 general coupling, but can shed light on the dynamics with
 analytical expressions.
\subsection{the case of\ \ $k_0Q_0=\pi$}
 For slow atomic center-of-mass motion, we start by
assuming that the atom is located at a position $Q_0$ such that
$k_0Q_0=\pi$ where $k_0=\omega/c$. We would like to emphasize that
$Q_0=\pi/k_0$ is a node of the sine mode function of the cavity
field, hence this position will not correspond to a potential
minimum of the cavity. This indicates that an additional  trap has
to be introduced to locate the atom. Alternatively, the cavity  can
be tuned to be blue detuned from the atomic transition and the mean
photon number in the cavity can be chosen high enough, such that the
atom is stable at that position. We will not specify the trap and
only consider an ideal situation where cavity loss and atomic
spontaneous emission are ignored. The Hamiltonian is this case
reduces to,
\begin{eqnarray}
H_{\pi}&=&\hbar(\omega-\xi q)a^{\dag}a+\frac{p^2}{2m}+\frac 1
2m\omega_m^2q^2 +\frac{\hbar\Omega}2\sigma_z\nonumber\\
&-&\hbar g_{\pi}(a^{\dag}\sigma^-+h.c.)q,\label{hpi}
\end{eqnarray}
where $ g_{\pi}=g\sin(\frac \pi L q)/q.$  We consider an example
where the amplitude of the mirror vibration is much smaller than the
length of the cavity ($q_{max}\ll L$), the coupling constant
approximately becomes  $g_{\pi}\simeq g\frac \pi L.$ The coupling
between the moving mirror and the CQED system (atom plus cavity
field) lead to modifying the atom-cavity interaction. By introducing
a transformation $p \rightarrow p^{\prime}=p,$ $q\rightarrow
q^{\prime}=q-\hat{\pi}/m\omega_m^2$, the Hamiltonian $H_{\pi}$ can
be rewritten as
$$
H_{\pi}=\hbar\omega a^{\dagger}a+\frac{p^{\prime 2}}{2m}+\frac 1 2
m\omega_m^2q^{\prime
2}+\frac{\hbar\Omega}{2}\sigma_z-\frac{\hat{\pi}^2}{2m\omega_m^2}.
$$
Here the operator $\hat{\pi}$ was defined as  $\hat{\pi}=\hbar \xi
a^{\dag}a+\hbar g_{\pi}(a^{\dag}\sigma^-+h.c.).$ In the adiabatic
limit of the CQED system, namely when the CQED system changes slowly
with respect to the fast-varying vibration of the mirror, we can
solve the oscillation of the mirror with fixed CQED
variables\cite{note2}, and rewrite the Hamiltonian as,
\begin{eqnarray}
H^{eff}_{\pi}=\hbar\omega a^{\dag}a+\hbar\omega_m(n_m+\frac 1
2)+\frac{\hbar\Omega}2\sigma_z
-\frac{\hat{\pi}^2}{2m\omega_m^2},\label{hpie}
\end{eqnarray}
where $n_m$ denotes the phonon number of the vibration mirror. In
the derivation of Eq.(\ref{hpie}), the mirror was assumed in a Fock
state $|n_m\rangle$, satisfying
$a_m^{\dagger}a_m|n_m\rangle=n_m|n_m\rangle$ with
$a_m=1/\sqrt{2m\hbar\omega}(m\omega q^{\prime}+i p^{\prime}),$ and
$a_m^{\dagger}=(a_m)^{\dagger}.$ This treatment is a good
approximation when the coupling of mirror vibration to the CQED
system does not induce transitions among the Fock states
$|n_m\rangle$ with different phonon numbers, namely
($|\psi(t)\rangle$ denotes any state of the CQED system)
$$\left |\langle \psi(t)|\frac{\langle n_m|\hbar g_{\pi}(a^{\dagger}\sigma^-+h.c.)q
+\hbar\xi a^{\dagger}aq|m_m\rangle} {\hbar \omega_m(n_m-m_m)}
|\psi(t)\rangle \right |<<1.$$
 Note that $\{|n+1, g\rangle, |n,e\rangle\}$
($n$ is the photon number in the cavity, while $|e\rangle$ and
$|g\rangle$ represent the excited and ground state of the two-level
atom, respectively) form an invariant subspace for the effective
Hamiltonian Eq.(\ref{hpie}), we may diagonalize this Hamiltonian and
obtain the following eigenvalues and corresponding eigenstates,
respectively,
\begin{eqnarray}
E_{\pm}(n)=\frac{H_{11}+H_{22}}2 \pm\sqrt{\frac 1
4{(H_{11}-H_{22})^2}+H_{12}^2}
\end{eqnarray}
and
\begin{eqnarray}
|+\rangle_n=\left(
            \begin{array}{c}
              \cos\frac{\theta_n}2 \\
               \sin\frac{\theta_n}2 \\
            \end{array}
          \right),
|-\rangle_n=\left(
            \begin{array}{c}
              -\sin\frac{\theta_n}2 \\
               \cos\frac{\theta_n}2 \\
            \end{array}
          \right). \label{drs}
\end{eqnarray}
Here $\theta_n$ can be determined by
$\tan\theta_n=\frac{2H_{12}}{H_{11}-H_{22}},$ and
\begin{eqnarray}
H_{11}&=&\hbar\omega(n+1)-\frac{\hbar\Omega}2
-\frac {\hbar^2}{2m\omega_m^2}[\xi^2(n+1)^2+g_{\pi}^2(n+1)], \nonumber\\
H_{22}&=&\hbar\omega n+\frac{\hbar\Omega}2-\frac {\hbar^2}{2m\omega_m^2}[\xi^2n^2+g_{\pi}^2(n+1)],\nonumber\\
H_{12}&=&-\frac{\hbar^2
g_{\pi}\xi}{2m\omega_m^2}(2n+1)\sqrt{n+1}=H_{21}.
\end{eqnarray}
The dressed states in Eq.(\ref{drs}) are different form that in the
Jaynes-Cummings (JC) model: the energy splitting depends on the
cavity field (through the photon number) more dramatically  than
that given by the JC model. As a consequence, the collapse and
revivals in the JC model would be modified, leading these
interesting features to disappear in this system.
\begin{figure}
\includegraphics*[width=0.9\columnwidth,
height=0.6\columnwidth]{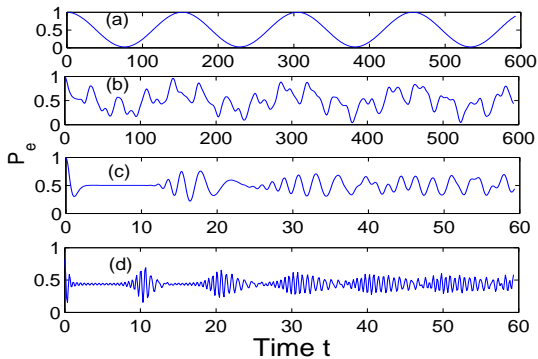}\caption{Population of the
atom in its excited state as a function of time. The cavity field
initially is prepared in a coherent state $|\alpha\rangle$, while
the atom in $|e\rangle$. $m=10^{-9}$ Kg, $L=5\times 10^{-3}$m,
$\omega_m=(2\pi)2.5\times 10^{3}$Hz,
$g=\omega=\Omega=(2\pi)4.5\times 10^{6}$Hz. The energy was rescaled
by $\hbar\omega_0$ ($\omega_0=10^{12}$ Hz here), and the time was
rescaled accordingly. (a),(b),(c) and (d) correspond to $\alpha
=0,1,3$ and $5$, respectively. } \label{fig2}
\end{figure}
This can be found in Fig.\ref{fig2}, where we present numerical
simulations for the dynamics governed by $H_{\pi}^{eff}$. The atom
was initially prepared in its excited state $|e\rangle$, while the
cavity field was assumed a coherent state $|\alpha\rangle$. Further
simulations show that the collapse and revivals are enhanced  by
large $\alpha$. This feature can be understood as nonlinear
atom-field couplings induced by the vibrating mirror, which speed-up
the population transfer between the interval degrees of the atom.
\begin{figure}
\includegraphics*[width=0.9\columnwidth,
height=0.6\columnwidth]{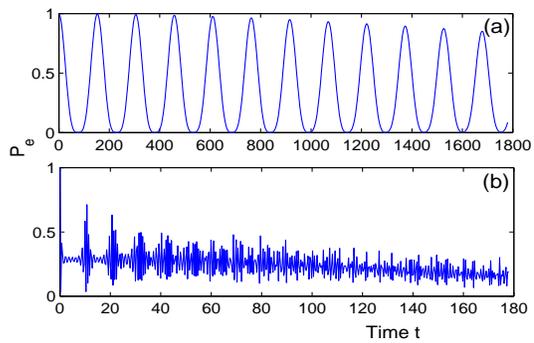}\caption{The same as
\ref{fig2}, but with  dissipation effects. $\Gamma=10g$, and
$\kappa=2g$ were chosen for this plot.  (a) and (b) are for
different $\alpha$, (a) $\alpha=0$, (b)$\alpha=5$.} \label{fig6}
\end{figure}
Before closing this subsection,  we briefly discuss the effect of
the atomic spontaneous emission and the cavity decay on the dynamics
of the system. The atomic spontaneous emission and the cavity decay
may be taken into account by adding an imaginary frequency shift
$i\Gamma/2$ to the Rabi frequency $\Omega$, and $i\kappa$ to the
cavity frequency $\omega$. The population of the atom in its excited
state as a function of time is shown in Fig. \ref{fig6}. Two
observations can be made from Fig. \ref{fig6}. (1) The atomic
spontaneous emission and the cavity decay make the population a
damping function of time (see Fig.\ref{fig6}-(a)); (2)The
spontaneous emission and the cavity loss spoil the collapse and
revival in the dynamics (see Fig.\ref{fig6}-(b)).

\subsection{the case of\ \ $k_0Q_0=\pi/2$}
When the atom is placed in a position satisfying $k_0Q_0=\pi/2$, the
Hamiltonian takes the form,
\begin{eqnarray}
H_{\pi/2}&=&\hbar(\omega-\xi q)a^{\dag}a+\frac{p^2}{2m}+\frac 1
2m\omega_m^2q^2 +\frac{\hbar\Omega}2\sigma_z\nonumber\\
&+&\hbar g(a^{\dag}\sigma^-+h.c.)-\hbar
g_{\frac{\pi}2}q^2(a^{\dag}\sigma^-+h.c.),
\end{eqnarray}
where  $g_{\frac{\pi}2}\simeq \frac{\pi^2g} {8L}.$ Following the
same analysis, we find that the coupling of the mirror to the cavity
field induce a nonlinear Kerr effect\cite{gong08}. The effective
Hamiltonian for the CQED system can be expressed as,
\begin{equation}
H_{\pi/2}^{eff}=\hbar\omega a^{\dag}a +
\frac{\hbar\Omega}2\sigma_z+\hbar
g(a^{\dag}\sigma^-+h.c.)-\frac{\hbar^2\xi^2}{2m\omega_m^2}
(a^{\dag}a)^2.
\end{equation}
This is exactly the Hamiltonian that describes a two-level atom in a
cavity filled with a nonlinear Kerr medium.

\section{slow mirror vibration and atomic center-of-mass motion}
Optomechanics has attracted considerable attention in recent years
not only because of its possible technological use but also because
of the theoretical interests in understanding the quantum-classical
transition. It is believe that entanglement can act as a bridge
between the quantum and classical world. So far, entanglement has
been experimentally prepared and manipulated using microscopic
quantum systems such as photons, atoms, and
ions\cite{bouwmeester00,esteve03}, stationary entanglement between
an optical cavity field and a macroscopic vibrating mirror has been
theoretically \cite{vitali071}, possibility to entangle two
macroscopic vibrating mirrors has been explored\cite{hartmann08}. It
would be interesting to extend the radiation-pressure-induced
entanglement to atomic center-of-mass motion and the vibration of
the mirror. From a theoretical point of view, the extension is
interesting, because this entanglement is shared between a
microscopic (atomic COM motion) and a macroscopic object(vibrating
mirror), moreover the atom and the mirror are not directly coupled,
but  interact with each other through the cavity field.

\begin{figure}
\includegraphics*[width=0.9\columnwidth,
height=0.6\columnwidth]{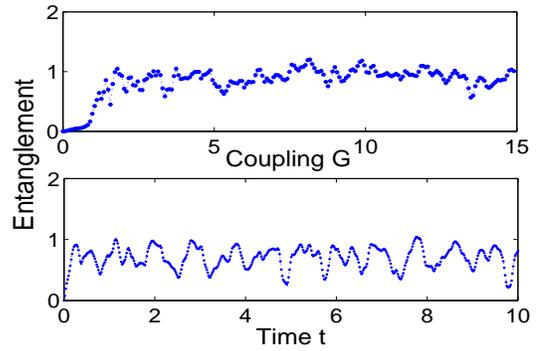}\caption{The entanglement
measured by the von Neumann entropy as a function of the coupling
constant $G$ (top) and time (bottom). $\omega=(2\pi)6\times 10^7MHz,
M=10^{-26}Kg, \Delta=10^3 Hz, \omega_m=(2\pi)8\times 10^7 Hz,
m=5\times 10^{-9}Kg, L=10^{-4}m$. The time was set in units of
$10^{-4}$s, and the coupling constant was plotted in units of MHz.
For the top panel $t=0.1$ms, and in the bottom panel $G=5\times
10^3$ Hz. The initial state chosen is $|1_c,1_b\rangle$, where
$|n_c,n_b\rangle$ denotes the Fock state of the system.}
\label{fig4}
\end{figure}

By considering the slow-varying mirror vibration and atomic COM
motion, we can solve the coupling of the cavity field to the atomic
internal degree of freedom in Eq.(\ref{h1}) first, with fixed $q$
and $Q$. This approximation is valid when the detunning
$(\omega-\Omega)$ is considerably large.  We denote
$|\phi_+,n\rangle$ and $|\phi_-,n\rangle$ the eigenstates, the
corresponding eigenvalues are given by,
\begin{widetext}
\begin{eqnarray}
U_{\pm,n}&=&\frac{2n+1}2(\hbar\omega-\hbar\xi q)
\pm\sqrt{\hbar^2g^2\sin^2(kQ)(n+1)+ \frac 1 4(\hbar\omega-\hbar\xi
q-\hbar\Omega)^2}.
\end{eqnarray}
\end{widetext}
Assuming the cavity field and the atomic internal degree of freedom
to follow an adiabatic evolution in state $|\phi_+,n\rangle$, we can
write the effective Hamiltonian for the mirror vibration and atomic
motion as
\begin{eqnarray}
H_{e}=\frac{p^2}{2m}+\frac 1
2m\omega_m^2q^2+\frac{P^2}{2M}+U_{+,n}.\label{effh}
\end{eqnarray}
This adiabatic treatment is valid if the coupling of the cavity
field to the atomic internal levels is far from resonance, such that
the level spacing $|U_{+,n}-U_{-,n}|$ is large, and population
transitions between $|\phi_+,n\rangle$ and $|\phi_-,n\rangle$ can be
ignored. Mathematically, this requires
$$\left |\frac{\langle\phi_+,n|\frac{\partial}{\partial Q}|\phi_-,n\rangle+
\langle\phi_+,n|\frac{\partial}{\partial
q}|\phi_-,n\rangle}{U_{+,n}-U_{-,n}}\right|<<1,$$ reminiscent of the
Born-Oppenheimer approximation.  In this situation, $U_{+,n}$ can be
approximated by,
\begin{eqnarray}
U_{+,n} &\simeq&-(n+1)\hbar\xi
q+\frac{\hbar g^2k^2Q^2(n+1)}{\Delta} \nonumber\\
&-&\hbar\xi q\frac{g^2k^2Q^2(n+1)}{\Delta^2},
\end{eqnarray}
where $\Delta=\omega-\Omega.$ In the remainder of this paper, we
shall choose $n=0$, implying no photon in the cavity while the atom
in its excited state. In this case, the effective Hamiltonian
Eq.(\ref{effh}) follows,
\begin{eqnarray}
H_e&=&\frac{p^2}{2m}+\frac 1 2m\omega_m^2
q^2+\frac{P^2}{2M}+\frac {\hbar g^2k^2Q^2}{\Delta}\nonumber\\
&-&\hbar \xi q -\hbar\xi q\frac{g^2 k^2Q^2}{\Delta^2}.
\end{eqnarray}
By the canonical quantization, we let $p,q,P, Q$ be operators, which
obey the commutation relations, $[q,p]=[Q,P]=i\hbar$, and the
others$=0.$ In terms of creation and annihilation operators, the
effective Hamiltonian reads,
\begin{eqnarray}
H_e&=&\hbar\omega_m(c^{\dag}c+\frac 1
2)+\hbar\omega'(b^{\dag}b+\frac
1 2)\nonumber\\
&-&\hbar G(c^{\dag}+c)(b^{\dag}+b)^2,
\end{eqnarray}
where the term $\hbar\xi q$ has been ignored, and
\begin{eqnarray}
\omega'\equiv gk\sqrt{\frac{2\hbar}{M\Delta}}, \ \ \ G=\frac{\xi
gk\hbar}{4\sqrt{mM\omega\Delta^3}}.
\end{eqnarray}
 Fig.\ref{fig4} shows the entanglement shared between
vibration of the mirror and the atomic COM motion, measured by the
von Neumann entropy.
Remarkably, the entanglement is always not zero when the coupling
constant $G$ is considerably large. And to some extend, we can say
the entanglement is insensitive to the coupling constant. For the
cavity with finite photon number , i.e., $n\neq 0$, the coupling
constant $G$ is proportional to $\sqrt{n+1}$, hence it increases the
strength of the coupling, however, it does not increase the
entanglement shared between the atomic COM motion and the vibration
of the mirror, as shown in Fig.\ref{fig4} (see the top panel). For a
specific coupling (for example,the effective coupling constant
$G=5\times10^3$Hz), the entanglement oscillates with time as shown
in the bottom panel of Fig.\ref{fig4}. It is interesting to study
when $\omega_m=2\omega^{\prime}$. In this case, the Hamiltonian
under the rotating-wave approximation is
$H_e=\hbar\omega_m(c^{\dag}c+\frac 1 2)+\hbar\omega'(b^{\dag}b+\frac
1 2) -\hbar G(c^{\dag}b^2+h.c.).$ The entanglement is a periodic
function of time in this case.
\begin{figure}
\includegraphics*[width=0.9\columnwidth,
height=0.6\columnwidth]{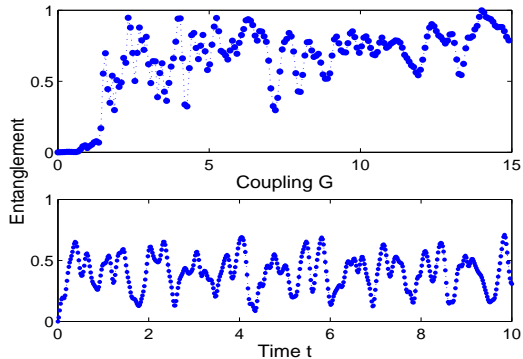}\caption{The same as Fig.
\ref{fig4}, but with different initial states. Here the initial
state of the vibration of the mirror is  chosen to be a thermal
state with $\beta=1/k_BT=10^{14}$Hz, while the initial state of the
atomic COM is $|1_c\rangle$.} \label{fig5}
\end{figure}
In order to study the entanglement  at finite temperature, we take a
thermal distribution of phonon $\rho_T=\sum_{n_b} e^{-\beta
n_b\omega_m}|n_b\rangle\langle n_b|/Z$ as an initial state, where
$Z=\sum_{n_b}e^{-\beta n_b\omega_m}.$  Numerical calculation for the
entanglement as a function of time and the coupling constant $G$ is
presented in Fig.\ref{fig5}. We find that the thermal effect spoils
the creation of entanglement. Nevertheless,  the entanglement can
still be created between the atomic COM motion and the vibration of
the mirror at room temperature. This is possible to observe  by
recent technology  that the radiation-pressure induced correlations
between two optical beams was demonstrated \cite{verlot09}.

\section{conclusion}
In conclusion, the dynamics and entanglement of an atom trapped in a
cavity with a movable mirror is studied. The key results come from
the atom-field-mirror coupling, which depends on the position of the
atom and the vibration of the mirror. By manipulating the detunning
between the atomic resonance  and the cavity field, the coupled
system falls into two remarkable regimes: fast CQED regime and slow
CQED regime. In the fast CQED regime, we have investigated the
dynamics of the CQED system, showing that the interesting feature of
collapse and revivals is significantly modified. The entanglement in
the atomic center-of-mass motion and the vibration of the mirror has
been studied in the slow CQED regime. Interestingly, we found the
entanglement is stable against the fluctuation of coupling. The time
evolution of the entanglement seems chaotic, except the special case
when $2\omega^{\prime}=\omega_m$.

\section*{ACKNOWLEDGEMENTS} This work was supported by  NSF of China
under grant  No. 10775023.


\begin{references}
\bibitem{note2} This is exact when the operator $\hat{\pi}$ defined
by $\hat{\pi}=\hbar\xi a^{\dagger}a+\hbar
g_{\pi}(a^{\dagger}\sigma^-+h.c.)$ is conserved in the dynamics (or
a constant of motion).
\end{references}
\end{document}